\documentstyle[12pt]{article}
\begin{document}
\baselineskip=22pt plus 0.2pt minus 0.2pt
\lineskip=22pt plus 0.2pt minus 0.2pt
\begin{center}
 \large
THE DARK MATTER PROBLEM IN LIGHT OF\\
QUANTUM GRAVITY
\vspace*{0.5in}

T.\ Goldman$^{*}$, J.\ P\'erez--Mercader$^{\dagger *}$, Fred
Cooper $^{*}$ \\
 and Michael Martin Nieto$^{*}$

\vspace*{0.25in}

 $^{*}$Theoretical Division, Los Alamos National Laboratory, \\
Los Alamos, New Mexico 87545 \\

\vspace*{0.25in}

$^{\dagger}$Laboratorio de Astrof\'{\i}sica Espacial
y F\'{\i}sica Fundamental, \\
Apartado 50727\\
28080 Madrid\\

\vspace{.5in}

Submitted to {\bf Nature}

\vspace{1in}

ABSTRACT

\end{center}

We show how, by considering the cumulative effect of tiny quantum
gravitational fluctuations over very large distances, it may be
possible to: ($a$) reconcile nucleosynthesis bounds on the density
parameter of the Universe with the predictions of inflationary
cosmology, and ($b$) reproduce the inferred variation of the density
parameter with distance. Our calculation can be interpreted as a
computation of the contribution of quantum gravitational degrees of
freedom to the (local) energy density of the Universe.

\pagebreak

Study of the observed rotation curves of galaxies and their
non--Keplerian fall-off has led to the conclusion that large amounts of
non--luminous   matter (Dark Matter) must exist, with a relative
concentration that must increase as one moves away from the center of
the particular galaxy. Furthermore, dynamical analysis of binary
galaxies, groups of galaxies, clusters of galaxies and the structures
known at the largest scales, show a slow, but clear, inferred increase
in the ratio of mass to luminosity, $M/L$. This translates into a
corresponding trend for the contribution of each structure to the
inferred cosmological density parameter $\Omega=8\pi G\rho/(3H^2)$,
where $G$ is Newton's constant, $\rho$ is the average mass density in
the Universe, and $H$ is the Hubble constant.

Thus, larger amounts of dark matter are required as the
scale of the structure grows. (See Figure 1 and Refs. 1--4.)

In addition, inflationary scenarios of the Early Universe, which
solve (among others) the horizon and flatness problems, while
maintaining the   successes of the Big-Bang, predict that $\Omega=1$.
On the other hand, primordial nucleosynthesis of the light elements,
based on the standard model of cosmology and known nuclear and
particle physics, bounds the total baryonic density of the Universe
to the range $0.010\leq \Omega_B h^2 \leq 0.035$. Here $h$ is Hubble's
constant, in units of $100 \rm{\,Km \, s}^{-1} {\rm Mpc}^{-1}$, which
is somewhere between $0.4$ and $0.8$. Thus, one can at most account for
less than 10\% of the energy density required to understand the
dynamics of large scales. This is the Dark Matter (DM) Problem.

The above conclusions are based on classical general relativity and,
where applicable, its non-relativistic limit, Newtonian gravity. In
this letter we will study the impact that quantum--mechanical
corrections to classical general relativity have on the DM problem. We
will find that quantum gravity effects can account for a major portion
of the effects attributed to DM and can also predict the value of
$\Omega$ as a function of distance scale over a $wide$ range of
distances differing by more than 25 orders of magnitude, without having
to introduce what is conventionally known as ``Dark Matter".

In quantum field theory, quantum effects lead to modifications of the
parameters of the theory due to the $unavoidable$ presence of virtual
processes. These modify the observed parameters and convert them into
scale-dependent quantities. In general, the scale dependence of the
corrections is logarithmic and their sizes are small, but if the range
of distances involved is very large, then their effects can build up
and become sizeable. For example, in Quantum Electrodynamics, QED
vacuum polarization effects increase the value of $\alpha_{em}$ (the
fine structure ``constant") from $\sim 1/137$ at atomic distances
to$^{\cite{goldmanross}} \sim 1/128.5$ at scales of the order of the
Compton wavelength of the $Z^{0}$--particle ($2\times 10^{-16}$ cm).
Similar corrections for QCD (the theory of the strong interactions)
make $\alpha_{strong}$ decrease from $\sim 0.179$, measured at the mass
scale of the bottom quark (Compton wavelength $4\times 10^{-15}$ cm) to
$\sim 0.101$ at the scale of the $Z^{0}$. As is well known, these
effects are predicted by the solutions to the renormalization group
equations (RGE) for $\alpha_{em}$ and $\alpha_{strong}$, and find
spectacular confirmation in high energy paricle physics experiments at
SLAC, Fermilab and CERN.

These effects may be understood by a classical analogy using fluid
dynamics. It is well known that for the same applied force, a longer
ship achieves a higher limiting speed than a shorter ship of the same
(mass and) cross-section presented to the water. Thus the force per
unit area exerted by the fluid depends on the length scale of the
``probe'' used to measure it. Physically this occurs because the longer
vessel reduces excitation of short wavelength modes of the fluid
that would ``close'' behind the shorter ship. In the quantum case,
coupling is to virtual modes for which the system lacks sufficient
energy to produce a real excitation. The strength of this coupling
still depends on the congruity between the scales of the modes and
the probe, thus producing a scale dependence to almost all measured
quantities.

We have asked the equivalent scale--dependence question for gravity and
investigated how such effects would manifest themselves. In order to
do that we start with the higher derivative theory of gravity described
by the following action:

$$
S = \int d^{4}x \sqrt{g} \left[ \Lambda - \frac{R}{16 \pi G} + a
W + \frac{1}{3} b R^{2} + \alpha_{V} R^{*} R^{*} + \kappa
D^{2}R\right] +  S_{surface} + S_{matter} ,
\eqno(1)
$$

\noindent
which  has been studied  in detail by, among others, Fradkin and
Tseytlin$^{\cite{fradkintseytlin}}$.

In this action, the $R^2$--terms may be thought of as those necessary,
at the classical level, to make the Einstein action (first two terms in
Eqn.(1)) quantum--mechanically better-defined in regimes where the
curvature becomes stronger. One obtains this action by expanding
certain more complete actions in a functional Taylor series in
derivatives of the metric.  Furthermore, we expect that the physics
described by this (or any reasonable) action to be independent of the
scale at which one is performing the experiments and defining the
physical quantities.  Because of the $R^2$--terms, the action is
renormalizable and one can apply these arguments$^{\cite{stelle}}$.

In fact, one can construct a ``Wilsonian'' version$^{\cite{wilson}}$ of
the action by substituting for the parameters appearing in the action
their scale--dependent (so-called ``running'' or effective) equivalents,
i.e., by substituting the solutions to the appropriate (one--loop in our
case) renormalization group equations computed from this action. This
Wilsonian action gives rise to ``classical'' equations of motion which
contain the effects of the quantum corrections to the physics at each
given scale. These corrections are incorporated in the values that the
effective parameters (including Newton's constant) take at each
distance scale. Thus, if we are interested in measuring Newton's
constant at some distance, $r$, we must include the effect that
(quantum) vacuum fluctuations have on the (micro) local curvature of
space--time. These fluctuations produce ripples in space--time
noticeable at distances of $O(10^{-33})$~cm. The ripples will be
conspicuous at these distances, at which they can be ``seen'' as
changes in the curvature of space--time or (equivalently) as a
modification in the value of Newton's constant. As we go past these
very short distances, the fluctuations occurring at the shortest
distances will still be felt and can be taken into account by using a
value of Newton's constant different from the one we used at the
smaller scale.  This is due to the fact that, unlike the case of
electric charge, there is no screening mechanism for gravity, and that
gravitation couples coherently to matter. Depending on how the
(unavoidable) quantum fluctuations of the geometry affect the suitably
averaged local curvature, the value at larger distances of Newton's
constant will grow with scale, decrease, or even oscillate!  Repeated
application of this procedure (very similar to the block
renormalization procedure in statistical mechanics$^{\cite{block}}$)
leads to an effective Newton's constant which is a function of $r$, but
which contains the effects due to those quantum gravitational
fluctuations for which the correlation length is not larger than the
length scale we are probing.

This variation with distance of $G$, or of any of the parameters
entering in the lagrangian, is controlled by the singular
behavior\footnote{To see the origin of this statement, consult, e.g.,
{\sl Physical Kinetics}, Vol. 10 of the Landau and Lifschitz's Course
of Theoretical Physics, Pergamon Press, New York.} of the appropriate
correlation function which serves to define that parameter. For $G$,
this may be obtained by computing the logarithmic divergences that
quantum fluctuations induce on the $R$--term of the action. Carrying
out this procedure for each of the parameters in the action of Eqn.
(1), one arrives at a set of renormalization group equations valid for
mass scales of less than $10^{-5}$~eV~c$^{-2}$ corresponding to lengths
longer than 1~cm. Here we have ignored photon and, possibly, neutrino
contributions.  The mass (and quantum mechanically corresponding
length) scale is conservatively chosen to avoid including (the effects
of) various additional degrees of freedom, which are expected to be
very massive and which are required for controlling the renormalization
behavior of the full theory at much shorter distance scales.

The effective parameters are given by the following
solutions$^{\cite{ourplb}}$:

$$
a(t) = a_{0} + \frac{133}{10} t ,
\eqno(2a)
$$

$$
\omega (t) = \frac{\omega_{+} - \alpha \omega_{-
} \left(\frac{a}{a_{0}}\right)^{-\gamma}}{1 - \alpha
\left(\frac{a}{a_{0}}\right) ^{-\gamma}} ,
\eqno(2b)
$$

$$
\frac{G(t)}{G_{0}} = \left[ \frac{\omega_{0}}{\omega(t)}
\right]^{\alpha_0}  \left[\frac{\omega(t) - \omega_{+}}{\omega_{0} -
\omega_{+}} \right]^{\alpha_{+}+\alpha_{-}}
\left[\frac{a(t)}{a_{0}} \right]^{\delta} .
\eqno(2c) $$

\noindent
Here $t \equiv (32\pi^{2})^{-1} log(\frac{\mu^2}{\mu_0^{2}})=(32
\pi^2)^{-1}\,\log(r_0^2/r^2)$; $\omega_{\pm} = \frac{-549 \pm 7
\sqrt{6049}}{200}$ are the two fixed points of $\omega(t)$; $\alpha
\equiv (\omega_{0} - \omega_{+})/(\omega_{0} - \omega_{-})$; $\gamma
\equiv\frac{100}{399} ( \omega_{+} - \omega_{-})\sim 1.36448$;
$\alpha_0=-13/5$; $\alpha_+ + \alpha_- =-18/5$ and $\delta=-1.24253$.
The quantity $\mu_0$ (or $r_0$) is a reference $momentum$ (or
$distance$) scale  at which the parameter is matched to the experiment:
$a(r_0)=a_0$, $\omega(r_0)=\omega_0$ and $G(r_0)=G_0$. The behavior of
the effective couplings $a(t)$ and $\omega(t)$ for $a_0 = 5.67$ and for
different initial values $\omega_0$ are shown in Figure 2.

The parameters $a$ and $b$ ($\equiv -\omega a$) are constrained by
laboratory and geophysical experiments$^{\cite{physrep}}$ as well as by
precise measurements\footnote{Notice that the usual PPN
approximation$^{\cite{ppn}}$ does not apply here$^{\cite{nonppn}}$,
since the effective potential, Eqn. (3) below, has finite range terms.}
of the orbital precision of Mercury. For the gravitational potential
due to a mass $m$, obtained from the static piece of the graviton
propagator, one finds in configuration space

$$
V(r)=\frac{-G(r)m}{r}\left[ 1-\frac{4}{3} e^{-m_2(r) r}+
\frac{1}{3} e^{-m_0(r) r}\right] ,
\eqno(3)
$$

\noindent
where $m_2^2(r)=(16 \pi G(r) a(r))^{-1}$ and $m_0^2(r)=(16 \pi G(r)
\omega(r) a(r))^{-1}$. From these expressions one obtains the following
formula for Newton's constant measurable in a Cavendish type
experiment:

$$
G_N(r)=G(r)
\left\{ \left( {1-{{\partial \ln G(r)} \over {\partial
\ln r}}} \right)\left[ {1+\sum\limits_{k=0}^1 {\gamma
_{2k}e^{-r/\lambda _{2k}}\left( {1+r/\lambda _{2k}} \right)}}
\right]
\right.
$$
$$
\left.
-\sum\limits_{k=0}^1 {\gamma _{2k}e^{-r/\lambda _{2k}}{r \over
{\lambda _{2k}}}{{\partial \ln f_{2k}} \over {\partial \ln r}}}
\right\} ,
\eqno(4)
$$

\noindent
where $f_0=\omega(r) a(r)$; $f_2=a(r)$; $\lambda_0=\hbar/(m_0(r) c)$;
$\lambda_2=\hbar/(m_2(r) c)$; $\alpha_0=+1/3$ and $\alpha_2=-4/3$.
Clearly, for values of $\omega(r)$ and $a(r)$ $\leq O(1)$, the
contribution from the Yukawa (exponential) terms is negligible at
distances of a few times Planck's length ($\sim 10^{-33}$cm). Thus, we
can safely ignore their contribution to the effective value of Newton's
constant.

Another point to bear in mind is that the gravitational degrees of
freedom contained in this action beyond the spin--2 graviton, and which
guarantee renormalizability, could at first sight lead to problematic
behaviors$^{\cite{hawkingghost}}$. The presence of a spin--2 massive
ghost (affectionately known as a ``poltergeist'') with mass $m_{2}$ can
lead to loss of unitarity\footnote{It is worth mentioning that some
quantum gravitational processes are known to violate {\it naive}
unitarity. One such process is the Hawking evaporation of a black
hole$^{\cite{strominger}}$. The correct implementation of unitarity in
quantum gravity remains an open question.} in both the high and low
energy regions.  In addition, the massive scalar could mean the
presence of an instability in the vacuum. The resolution of these
issues lies beyond the classical theory, in that, to correctly
understand their import, one has to consider the effects of quantum
corrections. These can come to the rescue since they significantly
affect the energy dependence of the mass which now has to be promoted
into a scale--dependent effective mass.

{}From the analytic expression for the mass of the poltergeist, one
sees that the attractive characteristic of gravity ($G >0$) makes it
into a bona fide, real mass, for $a(r)>0$. Furthermore, in the infrared
$a$ goes to zero, $G(r)$ increases only mildly, and $m_{2}^{2}
\rightarrow \infty$, so the effects of the poltergeist are $strongly$
suppressed. In the ultraviolet, where the problems from unitarity
become acute, quantum corrections also come to the rescue. For gravity
in its asymptotically free regime (i.e., for parameters such that
$G(r) \rightarrow 0$ as $r \rightarrow 0$), $m_2^2$ grows $faster$ than
$1/r^2$ and the effects of the poltergeist are subdominant at high
momentum transfer.

The remaining extra mode, the scalar field with mass $m_{0}$, will be
tachyonic if $m_{0}^{2}$ is negative. In ordinary quantum field
theories this implies that, at the length scales on which the scalar
field is active, the vacuum around which the quantum field theory is
built does not correspond to the lowest energy state for the system.
Here the same arguments we used for the poltergeist can be applied,
decoupling the tachyon in both the infrared and the ultraviolet and
thus rendering the tachyon harmless in both regions. In addition to
needing asymptotic freedom in $G$ for a correct behavior in the
ultraviolet, one also needs to guarantee that $\omega(r)$ remains
finite. This is readily seen to be true from the renormalization group
equation for $\omega(r)$. In the intermediate scale region the coupling
to matter fields can change the sign of $m_{0}^{2}$ and bring the
system to its true vacuum state. After the
decoupling$^{\cite{decoupling}}$ of the matter fields takes place, the
infrared regime sets in and the decrease of $a(r)$ at large $r$
decouples the scalar from the effective low energy physics.

{}From the above discussion, we conclude that we must stay in the
regions of parameter space where $a >0$ to have unitarity, whereas
$\omega$ can assume both positive and negative values.

Applying Wilsonian arguments to the basic quantum gravitational
action of \linebreak
Eqn.~(1), under the ansatz of a homogeneous and isotropic metric, one
finds that, for comoving distances greater than 1~cm, the energy
density parameter for the associated matter-dominated Universe,
$\Omega$, is promoted into a scale-dependent quantity $\Omega(r)$,
given by:

$$
\Omega(r) = \frac{8 \pi}{3 H^{2}_{0}} G_N(r) \rho_{m}.
\eqno(5)
$$

\noindent
(Because of the decoupling theorem$^{\cite{decoupling}}$, charge
screening and the distance scales we are considering, $\rho_{m}$ is
unaffected by quantum corrections at the one loop level).

In this expression $G_{N} (r)$ is given to a very high accuracy by

$$
G_{N}(r)=G(r) \left[ 1-\frac{{\partial \ln G(r)}}{{\partial \ln
r}} \right].
\eqno(6)
$$

We can write Eqn. (5) in the following reparametrization

$$
\Omega = \frac{8 \pi}{3 H^{2}_{0}} G_{lab} \delta(r,r_0)\rho_{m} ,
\eqno(7)
$$

\noindent
where $\delta(r,r_0)=G_{N}(r)/G_{N}(r_{0})$ and $r_0$ is the reference
laboratory distance at which G$_{lab}$ is measured.

General quantum field theoretical reasoning applied to gravity and to
the basic equations of cosmology thus leads one to the conclusion that,
to compare the inferred values of the density parameter for different
(size) structures in the Universe with the corresponding values
predicted by theory, one must use the ``improved'' expression given in
Eqn. (7). By construction, this takes into account the unavoidable,
scale-dependent effects of ``microscopic'' quantum fluctuations in the
geometry. The existence of a region in parameter space $(a_0,\omega_0)$
where the function $\delta(r,r_0)$ is a growing function of $r$,
together with Eqn. (7), opens up the possibility of understanding the
observational data on $\Omega$ over a wide range of scales and
reconciles the inflationary Universe scenarios with observation. In
fact, within a very ample region of parameter space, one can reach this
agreement without having to introduce any more contributions to
$\rho_m$ than what is consistent with the primordial nucleosynthesis of
the light elements. This means that we {\it need not} add any extra,
exotic contributions to the fraction of the critical energy density,
and thus do not need what is commonly known as dark matter to
understand the successes of the inflationary scenario.

The fraction of the critical energy density is constrained by
nucleosynthesis data to be$^{\cite {steigman}}$ $0.017 \leq \Omega_B
\leq 0.1$ or$^{\cite {white}}$ $0.02 \leq \Omega_B \leq 0.22$. We
saturate the nucleosynthesis upper bound (which we set at $\Omega_B
\leq 0.16$) and produce the fit shown by the continuous curve of Fig.
3, which is a plot of the function $\delta(r,r_0)$, for $r_0=1$ cm. At
once we see that for most of the typical systems, the inferred values
of $\Omega(r={\rm Typical\,\,\,Size})$ fall below our predicted
values.  This means that our scheme allows less mass to be contained in
these structures than what is commonly inferred for them by dynamical,
geometrical and (certainly) luminosity methods.\footnote{Other
independent tests are the ones based on the age of the Universe
combined with Geometry. These $(i)$ assume the validity of the Standard
Model and $(ii)$ must necessarily rely on $H$, whose value depends on a
variety of astrophysical assumptions. Such tests will be more
meaningful once purely geometrical measurements are available from, for
example, Hipparcos, the parallax measuring ESA satellite. We also see
that quantum gravitational effects incorporated via the effective
couplings provide a natural framework for $understanding$ the different
values of $\Omega(r)$ observationally inferred for the various typical
structures.}

Finally, it should be emphasized that our calculations take as a
starting point the model of quantum gravity of Eqn. (1) which, although
minimal, is nothing but one model in a class of models. It is
conceivable that other models\footnote{Notice that in the present
model, and for the chosen parameter values, $G$ changes by less than a
factor of 5 over 28 decades of distance. This perturbative leading
logarithm change should be compared with the equivalent change in grand
unified theories, where the effective couplings change (e.g. the strong
coupling constant) by a little more than a factor of 7 in 14 decades in
distance$^{\cite {goldmanross}}$.  This more gentle growth for the
gravitational coupling is consistent with the intuitive notion of
gravity as a much weaker force than the others.} yield different
behaviors for $\Omega(r)$. In addition, we have not included the
effects on the cosmological constant which, following the usual
practice, we set equal to zero. However, preliminary results in this
problem$^{\cite{vdwprl}}$, along the lines of what has been described
in this paper, show that for the model of Eqn. (1) the same mechanism
that makes $G$ grow with distance has the potential to produce an
exponentially strong suppression of the physical value of the
cosmological constant from its value at scales corresponding to much
less than $10^{-5}$~eV~c$^{-2}$ to the largest known distances, of the
order of $10^6$~kpc. This adds weight to our contention that many
issues of large scale astrophysics, not just Dark Matter, require the
inclusion of quantum mechanical effects on gravitation before they can
be properly evaluated and resolved.

\pagebreak

\pagebreak
\noindent
{\large Figure Captions}

\vspace*{.2in}

\noindent
Fig.\ 1 \hspace*{0.1in} Plot of the inferred density parameter
$\Omega_0$ for different typical structures. This plot is an update of
the one given in Reference {\cite{trend1}}. The data plotted are
extracted from that given by References {\cite{trend2}},
{\cite{trend3}} and {\cite{white}}. The different structures are: {\it
(A)}, Solar Stellar Neighborhood; {\it (B)}, Visible Galaxies; {\it
(C)}, Extended Rotation Curves; {\it (D)}, Binary Galaxies; {\it (E)},
Galaxy Groups; {\it (F)}, Rich Clusters; {\it (G)}, Virgo Infall; {\it
(H)}, IRAS Galaxies.

\vspace*{.1in}

\noindent
Fig.\ 2 \hspace*{0.1in} The scale evolution of the effective couplings
$a(r)$ and $\omega(r)$ in the renormalizable theory of gravity
considered in the text and described by Equation (1). The running of
the coupling $\omega$ is shown for different initial values. Lines
showing the two fixed points, $\omega_{+}$ and $\omega_{-}$, are also
drawn.

\vspace*{.1in}

\noindent
Fig.\ 3 \hspace*{0.1in}  The product of the ratio of the
nucleosynthesis upper bound on the (baryonic) matter density to the
critical matter density with the ratio of the scale dependent value of
Newton's constant to the laboratory value, as a function of scale, for
a point mass, compared with data on the value of the deceleration
parameter inferred using the laboratory value of Newton's constant.
{}From $10^{-7} kpc$ down to laboratory scales ($3 \times 10^{-22} kpc$),
the curve is visually indistinguishable from a horizontal straight
line. A horizontal line marks the nucleosynthesis upper bound on
the contribution of baryonic matter to the critical density.

\vspace*{.3in}

\end{document}